\documentstyle[11pt]{article}

\makeatletter
\def\dif{{\rm d}}
\def\deriv{\@ifnextchar[{\@deriv}{\@deriv[]}}
   \def\@deriv[#1]#2#3{\mathchoice%
{{\dif^{#1}#2\over\dif{#3}^{#1}}}{{\dif^{#1}#2/\dif{#3}^{#1}}}%
{{\dif^{#1}#2\over\dif{#3}^{#1}}}{{\dif^{#1}#2/\dif{#3}^{#1}}}}

\def\secteqno{\@addtoreset{equation}{section}%
\def\theequation{\thesection.\arabic{equation}}}
\newcounter{subequation}
\def\thesubequation{\alph{subequation}}
\def\sneqnarray{\stepcounter{equation}\let\@currentlabel=\theequation
\setcounter{subequation}{1}
\def\@eqnnum{{\rm (\theequation.\thesubequation)}}
\global\@eqcnt\z@\tabskip\@centering\let\\=\@eqncr\let\@@eqncr=\@@sneqncr
$$\halign to \displaywidth\bgroup\@eqnsel\hskip\@centering
 $\displaystyle\tabskip\z@{##}$&\global\@eqcnt\@ne
 \hskip 2\arraycolsep \hfil${##}$\hfil
 &\global\@eqcnt\tw@ \hskip 2\arraycolsep $\displaystyle\tabskip\z@{##}$\hfil
  \tabskip\@centering&\llap{##}\tabskip\z@\cr}
\def\endsneqnarray{\@@sneqncr\egroup $$\global\@ignoretrue}
\def\@@sneqncr{\let\@tempa\relax
   \ifcase\@eqcnt \def\@tempa{& & &}\or \def\@tempa{& &}
   \else \def\@tempa{&}\fi
     \@tempa \if@eqnsw\@eqnnum\stepcounter{subequation}\fi
     \global\@eqnswtrue\global\@eqcnt\z@\cr}
\def\nobiblabels{\def\@lbibitem[##1]##2{\@bibitem{##2}}}
\makeatother
\def\tabaddress#1{{\small\it\begin{tabular}[t]{c}#1\\[1.2ex]\end{tabular}}}

\def\ben{\begin{enumerate}}
\def\een{\end{enumerate}}
\def\beq{\begin{equation}}
\def\eeq{\end{equation}}
\def\bea{\begin{eqnarray}}
\def\eea{\end{eqnarray}}
\def\beann{\begin{eqnarray*}}
\def\eeann{\end{eqnarray*}}
\def\beasn{\begin{sneqnarray}}
\def\eeasn{\end{sneqnarray}}

\def\UBECM{Departament d'Estructura i Constituents de la Mat\`eria\\
   Universitat de Barcelona\\
   Av.~Diagonal 647\\
   08028 Barcelona\\
   Catalonia, Spain}

\let\oldGamma=\Gamma
\def\Gamma{{\bf\oldGamma}}

\title{Observables, gauge invariance, and the role of the 
observers in the limit from general relativity to special relativity}

\author{Josep M. Pons$^a$
\\
\tabaddress{$^a$\UBECM}
\\[2mm]
\small e-mail: 
{\tt pons@ecm.ub.es}
}

\date{\today}

\begin{document}

\maketitle

\begin{abstract}

Some conceptual issues concerning general invariant theories, with special 
emphasis on general relativity, are analyzed. The common assertion that 
observables must be required to be gauge invariant is examined in the light 
of the role played by a system of observers. Some features of the reduction 
of the gauge group are discussed, including the fact that in the process of 
a partial gauge fixing the reduction at the level of the gauge group and
the reduction at the level of the variational principle do not commute.
Distinctions between the mathematical and the physical concept of 
gauge symmetry are discussed and illustrated with examples. 
The limit from general relativity to special relativity is considered 
as an example of a gauge group reduction that is allowed in some specific
physical circumstances. Whether and when the Poincar\'e group 
must be considered as a residual gauge group will come 
out as a result of our analysis, that applies, in particular, to  
asymptotically flat spaces. 

\end{abstract}

\section{Introduction}

Poincar\'e invariance and general ---reparameterization, or diffeomorphism--- 
invariance are usually considered very separate conceptual fields. 
Thus, whereas the former, which is a rigid symmetry, is given
physical significance, the latter, which is a gauge symmetry, is 
considered unphysical. Gauge invariance is a symmetry of 
the equations of motion that depends upon arbitrary functions and their 
derivatives, so a given set of initial data can lead to different future
evolutions that will be related by gauge transformations. A deterministic
interpretation of the physical time evolution of the system then requires
these gauge related trajectories to be physically equivalent.
This is the case
for diffeomorphism invariance in general invariant theories. 

We always consider the gauge group as acting on the space of field 
configurations ---including, if necessary, particle-like world lines. 
In the particular case of a general invariant theory,
part of the gauge group ---or the whole of it in 
absence of other gauge symmetries--- is induced by the 
diffeomorphism group originally acting on the spacetime manifold. 
In this case an element of the gauge
group prescribes the action of a given diffeomorphism on a given field 
configuration. In General Relativity (GR) this means that 
different metrics, and different fields in general, can undergo different 
diffeomorphisms \cite{Pons:1996pr} \cite{Pons:1999xu} under a single element
of the gauge group. An example is, for instance, the action of the 
canonical generators of the gauge group
in Einstein-Yang-Mills theories, where their associated 
transformations exhibit a compulsory dependence on the lapse
and shift functions and on the time component of the gauge vector field 
\cite{sal/sunder/83a} \cite{sal/sunder/83b}. 
The gauge group in a general 
invariant theory is thus a diffeomorphism-induced gauge group.

Other cases of gauge invariance are associated
with {\it internal} groups of symmetry, as in Yang-Mills type theories. 
Instead of the ordinary conserved quantities associated 
---through Noether theorem--- with rigid symmetries, gauge invariance leads to 
constraints that restrict the physical phase space. For theories 
exhibiting this internal gauge invariance, such as electromagnetism 
within the framework of
Special Relativity (SR), the physical interpretation 
requires the observables to be 
gauge invariant ---at least on shell---.  Accordingly, in a general invariant 
theory, it seems natural that the observables be required to be 
invariant under diffeomorphisms. In contrast, no one will ask the observables
in a Poincar\'e invariant theory to be Poincar\'e invariant, that is, to 
commute with the generators of the Poincar\'e algebra. There is thus a sharp 
contrast between the concept of observables in a general invariant theory 
and in SR. 

General invariance is a basic principle for any physical theory because of the
universality of gravitation, that couples directly to matter and energy.
Wherever there is ``something'', there is gravity. Then the principle 
of equivalence, which we subscribe, will 
lead us to general invariance, that is, invariance under diffeomorphisms. 

General Relativity (GR) is a classical example of a general invariant theory. 
It is common to general invariant theories to admit formulation in any set 
of coordinates, their physical contents being
independent of the choice of the coordinates used to describe the system.
But then, how can we ever use and attach physical meaning to a theory like
SR, that is not general invariant? The answer is: as an approximation. 
The concept of a Poincar\'e invariant fundamental 
physical theory with formulation in Minkowski space is acceptable as long 
as the gravitational interaction is not relevant for the phenomena the 
theory is intended to describe. 
Otherwise the theory should be formulated in a general invariant way. 
To say it in another way: the reason why a given classical  
fundamental theory is not general invariant but only Poincar\'e invariant 
is because such a theory is only 
an approximation ---that may be an extraordinarily good approximation--- 
to a physical situation where it is reasonable to 
ignore gravity. Consequently, there should exist a procedure, 
applicable at least in some
definite physical situations, to go from a general invariant theory 
to a Poincar\'e invariant one. 
This is the meaning for GR to be a relativistic theory \cite{will}.

Let us see how this smooth transition from general invariance to Poincar\'e 
invariance can be performed. Consider that we start with a theory of GR with 
matter in a physical situation such that it is allowed to neglect the 
gravitational interaction when compared to
other, much stronger, interactions. In such cases we take the spacetime
manifold as non dynamical and, moreover, we make the approximation 
that it is flat. What remains then of the gauge 
invariance associated with the freedom of reparameterization? 
At this point nothing seems to have changed because general invariance is 
still present, but the dynamics has undergone
substantial changes for, being the gravitational field non dynamical, there
are no longer equations of motion associated with the metric of spacetime. 
In particular, the constraints corresponding to the momenta canonically 
conjugate to the lapse and shift functions disappear because lapse and
shift, as components of the metric, are no longer dynamical variables. They
are just background. Diffeomorphism invariance is still in place in the sense
that we are free to reparameterize, but there are no constraints associated
with this invariance because the metric has ceased to be dynamic. At this
point it may prove convenient to make a partial gauge fixing and to decide 
to work only with systems of coordinates such that the metric takes the 
Minkowski form. Notice the crucial fact that this gauge fixing is partial 
because some gauge transformations survive it: what is left from the 
diffeomorphism group after this gauge fixing is just the Poincar\'e group. 
Thus, the transition from a general invariant theory to a Poincar\'e 
invariant one satisfies an obvious consistency check, that of the reduction of
the gauge group.

According to this interpretation, the Poincar\'e group is a leftover, that is, 
a residual gauge group, that appears in a general invariant theory
as a byproduct of a) ignoring gravity (which must be sustained on physical 
grounds), and b) selecting the Minkowski form for the flat metric. 

However, this construction of a smooth limit 
from GR to SR meets with some conceptual obstacles.
  
One first difficulty appears when both theories, the original 
general invariant 
theory and the reduced Poincar\'e invariant theory, are formulated in 
variational terms; in this case, the limit from GR to SR 
contains two different processes that do not commute: the process of 
reducing the gauge
group, and the process of reducing the dynamics or, more properly, the 
process of reducing the variational principle. This issue will be explained 
and clarified in section 3.

A second difficulty stems form the very concept of observables.
If an observable in GR
is required to be diffeomorphism invariant, our limiting procedure to SR 
will convert it in a Poincar\'e --the residual gauge group-- invariant  
object, which is a concept very far away from the standard concept of 
observable in SR. 
So it seems as if there is a discontinuity in a limiting procedure from
GR to SR, at least concerning the concept of observables. Is there any way out?

Our answer is in the positive, because so far we have been considering 
observables without any regard to the observers. It is time to consider them, 
as the piece that mediates
between the observables and their physical interpretation. Let us have a 
look first to SR. In SR one defines 
privileged systems of observers: those sitting in the inertial reference 
frames. Once an inertial reference frame 
is selected, we fill it, as an idealization, of observers sitting each at any
value of the space coordinates, with clocks that are synchronized. 
Observers in SR are then labeled (with Cartesian space 
coordinates, for instance) in a way that, together with their 
synchronized clocks, allows for a direct physical interpretation of their 
readings in terms of lengths and lapses of time. All these readings are values
for the observables of length and time. 

In contrast, observers in GR usually have a very different status, for
it is not possible in general to attach direct physical significance to
the coordinates they are working with. Our aim is to look for a concept of
observers in GR that could 
eventually be consistent with the usual concept of observers in SR when 
the limiting procedure from GR to SR is taken.

One first approximation to observers in GR is that they do not interfere with
the geometry, that is, they are idealized observers playing only a kinematical
role, as test bodies. 
Let us examine, in the framework of GR, the basics for defining physical
length and physical time, consistent with a system of observers, and keeping
an eye to an eventual limiting procedure to SR (these considerations 
will be expanded with all technical details in the next section).
As we have said, our observers are
idealized: they must be able to gather information from their surroundings but
without affecting neither each other nor the spacetime manifold. 
Proper time is always defined for every observer --that follows 
a time-like world line. Since we are free to choose coordinates, it is
natural to make this choice in agreement with what we know of the observers 
in SR, so we choose coordinates that will be 
comoving with the observers, that is, the observers will sit at fixed values 
of the three-space coordinates. The distance from an observer to an 
infinitesimally close neighbor observer can be defined but it can not be 
extended, in general, to finite distances. In the next section we will review
the conditions set for these observers in order to be consistent 
with a transitive synchronization that is preserved in time. 
In such case, we say that the observers are privileged.

\vspace{4mm}

These privileged observers allow for the emergence of 
a physical concept (always associated with this system of observers):
that of equal time surfaces. We submit that this concept is 
``physical'' ---observable--- not because it is diffeomorphism 
invariant, which is not, but because it is associated with a given 
system of privileged observers, in the same way as we consider physical 
concepts in SR (for instance: simultaneity of two events 
is a physical observation, though it only makes 
sense when associated with a given SR inertial reference system). 

As long as a system of observers of this type can be set in a given general
invariant theory, events will be referred to it. 
In this sense, standing with a given system of observers 
is equivalent to performing a partial gauge fixing 
(technical details in the next section). 
Now the observables need not be given in a gauge invariant
(that is, diffeomorphism invariant) way but only as 
invariant under the residual gauge transformations that preserve the system 
of observers. This is the key point that will help to dissolve 
the conceptual difficulties we faced with the limit
GR $\rightarrow$ SR. Notice also that this limit does not need 
to be performed in a full spacetime manifold but it can be circumscribed 
to a specified region of it: the spacetime region where the experiences we 
try to describe within our theoretical framework are taking place. 
On the other hand, full gauge invariance of the observables may be 
restored by introducing a diffeomorphism invariant description of the 
observers into the formalism.

The concept of observability in GR has been discussed long ago 
\cite{berg61b, berg-book}. It has been deeply analyzed in 
\cite{Rovelli:1991ph}, where the introduction of 
the ``non-local'' and the ``local'' 
points of view ---based upon the non-existence
or existence of observers associated with a given reference system--- 
is very illuminating. In this 
sense we think that our approach keeps a close relationship with, and is 
indebted to, \cite{Rovelli:1991ph} in that some
eventual contradictions in the literature are solved by clarifying the role 
of observers. Our scope, though, is somewhat different, 
for we try to meet the conditions to produce a smooth transition from a 
general invariant theory to a Poncar\'e invariant theory.

The paper is organized as follows. In the next section we technically 
substantiate the discussion introduced above. In section 3 we make some
distinctions between the physical and the mathematical concept of
gauge transformations, and a subsection is devoted to a specific example 
featuring some of these ideas. Further applications of these ideas to 
asymptotically flat spacetimes are given in section 4. Finally 
section 5 is devoted to conclusions.


\section{Privileged systems of observers in GR}

 In GR, or in any general invariant 
theory, given (at least in the spacetime region of our interest) a 
three-parametric congruence of time-like world lines ---that is, a platform
\cite{soffel}---, 
we can always take coordinates in a 3+1 decomposition adapted to this  
congruence, such that each world line has a fixed value for the space 
three-coordinates (comoving coordinates). We can think of observers attached 
to these world lines. Then, by use of light rays one can define a 
synchronization between one observer and another infinitesimally close. 
The vector fields
that connect synchronized infinitesimally close neighbor observers are 
$$
{\bf X}_i := \partial_i - {g_{0i}\over g_{00}} \partial_0,    
$$
where $g_{\mu\nu}$ ($\mu,\nu = 0, \dots, 3;\ i = 1,2,3$) 
are the components of the spacetime Lorentzian metric. The
vectors ${\bf X}_i$ span the directions orthogonal to the vector field 
tangent to the world lines.
  
A good synchronization is characterized by two properties: first, 
it must be consistent with that of the neighbors' 
(transitive property for infinitesimally close observers: if $A$ is 
synchronized with $B$, and $B$ with $C$, then $A$ and $C$
must be synchronized); 
and second, it must be kept in time 
(to be a physical synchronization: If $A$ and $B$ are synchronized at time zero,
they must be so at later times).

\underline{Transitivity of the synchronization.}
The first property is equivalent to saying that the vector fields ${\bf X}_i$
form a distribution, that is, they locally define a three-surface. Due to the
particular form of ${\bf X}_i$, this property amounts to commutativity:
\beq
[{\bf X}_i, \, {\bf X}_j] = 0.
\label{commut}
\eeq

This property is equivalent to that of the irrotationallity of the world lines
of our observers.
A weaker sense of consistency can also be defined for a two-parametric family
of observers. For instance, the property
\beq
[{\bf X}_1, \, {\bf X}_2] = 0,
\label{commut12}
\eeq
already guarantees the existence of two-surfaces such that observers sitting
in those surfaces (a two-parametric family of observers) can be consistently 
(transitive property) synchronized.

\underline{Preservation of the synchronization.}
Now enters the second property for a good synchronization: preservation in time.
Neighbor observers sitting in $x^i$ and $x^i + \epsilon^i$, with 
$\epsilon^i$ some infinitesimal constants, will preserve
their synchronization (initially produced with the help of the vector field 
$\epsilon^i{\bf X}_i$)) along the time evolution if and only if
\beq
[\epsilon^i{\bf X}_i, \, {\bf X}_0] = 0,
\label{commut-i0}
\eeq
where ${\bf X}_0$ is the vector field generating the evolution in proper
time:
$${\bf X}_0 := {1 \over \sqrt{-g_{00}}} \partial_0.
$$

The interpretation of (\ref{commut-i0}) is obvious: displacement to the 
neighbor observer and evolution in proper-time commute. If the preservation 
in time of the synchronization is required for any set of $\epsilon^i$,
then the condition is: 
\beq
[{\bf X}_i, \, {\bf X}_0 ] = 0.
\label{commut0}
\eeq

Equations (\ref{commut}) and (\ref{commut0}) describe the full compatibility
of a system of observers with a synchronization that is preserved 
in time.

\underline{Geodesic condition.}
Up to now we have said nothing regarding the motion of our 
observers in spacetime. In this respect, we must take notice that equations 
(\ref{commut0}) are exactly the conditions for our observers to follow a 
geodesic of the metric \cite{wald84}. 
The motion of the privileged observers is free fall.

In conclusion, irrotationallity (\ref{commut}) and geodicity (\ref{commut0}) 
are the two conditions that our
world lines must satisfy in order to be associated with privileged observers
in GR. This corresponds to a platform that is locally proper-time 
Einstein-synchronizable \cite{soffel,sachs77}.

\vspace{5mm}

Equation (\ref{commut}) implies that we can make a change of coordinates
\beq
(x^i, x^0) \Rightarrow (x'^{i}, x'^{0})
\label{xtoxprime}
\eeq
such that ${\bf X}_i$ becomes  $ \partial'_i$ in the new coordinate system . 
This leads to $x'^i = x^i$ 
(the observers still sit in the same values of the three-coordinates) and to a 
function $x'^0(x^i, x^0)$, solution of the equations 
\beq
{\bf X}_i(x'^0) =0.
\label{Xonx0}
\eeq

The metric tensor is rewritten, under this change of coordinates, as
\beq
g'_{0j} = 0, \quad
g'_{ij} = g_{ij} - {g_{i0}g_{0j} \over g_{00}}, \quad
g'_{00} = g_{00}{1 \over ({\partial x'^{0} \over \partial x^{0}})^2}.
\label{newmetr}
\eeq

The interpretation of (\ref{newmetr}) is clear. According to the first 
equation, the lapse functions vanish in the new coordinate system. As regards
the second equation,  
$g_{ij} - {g_{i0}g_{0j}\over g_{00}}$
is just the Landau metric \cite{landau} that defines 
the spatial distance between 
infinitesimally close observers in comoving coordinates; since in the new 
coordinates these observers are synchronized, their spatial distances are
directly determined by the new 3-metric $g'_{ij}$. The last equation 
is a consistency requirement for the invariance of proper times as
computed in the old or in the new system of coordinates.

In the new system of coordinates, the condition  (\ref{commut0}) for
the preservation of the synchronization has a new interpretation: $g'_{00}$, 
when expressed in the new coordinates, must be a function of $x'^{0}$
exclusively. Then the proper time between  $x'^{0}_A$ and $x'^{0}_B$ 
is independent of the observer, that is, it 
will not depend on the three-coordinates, which means that synchronization will
be preserved in time. It is easy to verify that the requirement
$$
\partial'_i (g'_{00})=0
$$
is equivalent to (\ref{commut0}), provided, of course, the fulfillment 
of (\ref{commut}).

With the new coordinates the metric takes the form (we drop the primes)
\begin{eqnarray*}
    (g_{\mu\nu}) & = &
            \pmatrix{ g_{00}(x^0) & 0 \cr
                             0  & g_{ij}(x^0, x^i)}   \ ,
\end{eqnarray*}
and now, with a simple reparameterization of the time coordinate, we can get 
$g_{00} = -1$, which means that we are using the proper 
time of the synchronized observers as the new time coordinate.

Obviously, a system of observers in free fall is not always 
physically realizable. It is common in cosmology, when the observers 
can be idealized as comoving with the galaxies (up to the peculiar motions), 
but if 
we study the motion of a particle around a Schwarzchild solution of Einstein 
equations, it is not the kind of observers we will use. 
Notice, however, that in this last 
example gravity can not be neglected, and we are
interested in cases where a smooth transition from GR to SR can be taken.

\vspace{5mm}

Coordinates such that the metric satisfies 
\beq
g_{00}= -1 , \quad g_{0j}=0
\label{gauss}
\eeq
are known as Gaussian coordinates \cite{weyl,Ohanian:1994uu,wald84}. Obviously
a suitable gauge transformation ---a diffeomorphism--- can always set 
the coordinates in a local patch to be Gaussian.
Gaussian coordinates are associated therefore
to inertial (that is, moving along geodesics) comoving observers, that are 
synchronized to each other and such that 
their clocks give the time coordinate (proper time) for an event. 
These families of
observers associated with Gaussian coordinates (or coordinates that can be
put in Gaussian form by a change of the type (\ref{xtoxprime}))
are therefore privileged families of 
observers. This is as close as we can get in GR to the usual 
systems of observers of SR. 

Infinitesimal coordinate transformations 
$x^\mu \rightarrow x^\mu - \epsilon^\mu$ 
that keep $g_{00}=1$ and $g_{0j}=0$ must satisfy (the comma
denotes partial differentiation):
\beq
\epsilon^0_{,0} = 0 \ , \quad  
\epsilon^0_{,i} = g_{ij} \epsilon^j_{,0}.
\label{gaussrestr}
\eeq

Notice that (\ref{gaussrestr}) implies that ${\vec \epsilon}_{,0} = 0$ 
if and only if $\nabla \epsilon^0 =0$,
where ${\vec \epsilon} = (\epsilon^1\, ,\, \epsilon^2\, , \, 
\epsilon^3)$ and $\nabla \epsilon^0 = (\partial_1\epsilon^0,\,
\partial_2\epsilon^0,\, \partial_3\epsilon^0)$. So when, in addition to 
(\ref{gaussrestr}), we have 
$\nabla \epsilon^0 =0$, the transformation amounts to a relabeling
of the spatial coordinates for the observers and a rigid translation of the
zero of time: the system of observers remains the same. Instead, when
$\nabla \epsilon^0 \neq 0$, the change of spatial coordinates
involves the time and therefore the system of observers changes (these are
the ``boosts'' in GR). In order
to do physics it is legitimate to stay with a given system of observers and
to make a choice of the zero of time for their synchronized clocks. 
These choices amount to a partial gauge fixing. Under this interpretation,
physics does not require our observables to be invariant under
general diffeomorphisms, but only under diffeomorphisms that are 
compatible with the system of observers we are working with 
---but see the remarks below. In this sense, a distance as given by the 
three-metric is an observable (because it is invariant under the 
arbitrary three-space diffeomorphisms described by (\ref{gaussrestr}) when 
restricted to $\nabla \epsilon^0 =0$). This distance may change in 
time if $g_{ij}$ has a time dependence; this only means that such 
distance, always referred to our system of observers, is not 
time invariant, but it is still physical, observable.

This discussion suggests that the usual claim that, in general invariant 
theories, only gauge invariant
quantities, that is, diffeomorphism invariant quantities, are acceptable 
as observables, must be re-analyzed. Diffeomorphism invariance 
must be required if we lack of a system of observers of the type we 
have been building up, but not when
it is possible to set up a system of observers consistent 
with synchronization. Then we only need to ask the observables to be  
invariant under transformations that do not change the family 
of observers. Once said that, let us notice that the alternative, 
fully diffeomorphism invariant description, can always be restored 
by including the observer's world-lines in the physical description. 
See the remarks below.  

Summing up, we have cut the original set of gauge transformations along several
steps. In the first step, only gauge transformations preserving the Gaussian 
condition are allowed. In the second, the gauge group is further restricted to
the transformations that preserve the system of observers 
(three-diffeomorphisms
and proper time translations), and finally, by deciding the zero
of the time coordinate, we are left with the three-diffeomorphisms as the 
residual gauge freedom. The transformations that change the system of 
observers are interpreted as a change of description, from one system
of observers to another. These transformations have some analogy with the
boosts in SR.

Now consider that, in a given region of spacetime, a smooth transition 
to SR can be taken. Recall that with our privileged 
observers in GR we have already fixed the gauge freedom of time 
translations by deciding the zero of the time coordinate, now in SR 
we can even decide where to place the origin (the zero) of the 
three-coordinates and to choose three orthogonal
space directions to be labeled as the $x \, , y \, , z$ directions. All these 
choices are gauge fixings, and we 
can then agree that our descriptions of physical events will be referred to
this fixed framework. This means that we have finally shrunk the residual
gauge group to nothing and, therefore, there are no longer gauge 
invariance requirements for our observables. 
The result is that a simple position at a given
time, the third component of an angular momentum, etc., become 
observables. 
Thus observability increases a great deal when the limit is taken from 
GR to SR (by neglecting gravity when it is physically acceptable),
for Cartesian coordinates in the latter are endowed with a direct 
physical significance that has no parallel in the former. 
If in our previous discussion within GR we found that the relational 
concept of observables had room for systems of observers that will 
refer the physical events in a
three-space diffeomorphism invariant way, now in SR the system of observers,  
directly associated with a given Cartesian coordinate frame, carries
direct physical observable content,  
thus enlarging enormously the concept of observability. This explains why
the concepts of observability look so different when considered in GR or 
in SR.

Let us finally make three important remarks. 

\underline{First}, our observers can be conceived
as test bodies \cite{mir} forming a grid of sensors moving in free fall 
that, in an acceptable approximation for the phenomena 
they are set to describe, 
collect information from the media without producing  
or receiving any other (significant) perturbation in or from it 
\footnote{These idealized conditions are close to those introduced 
in section 3 of \cite{Rovelli:1991ph} for the material reference systems.}. 
Observers, thus, must be physical. Whether they account or not for some 
matter degrees of freedom depends on the approach we take for their 
description, as is clarified in the next remark. 
They are part of the -classical- physical system but 
with an ideal beauvoir -a non-interference condition- with respect 
to the rest of it. 
In this case, the relational concept \cite{Rovelli:1997qj} of observables in 
GR, which is the statement that objects are only localized with respect
to other objects and not with respect to background space, can take
advantage of this structure of observers added to the physical framework, 
and become a concept suited to perform the limit GR $\rightarrow$ SR. 

\underline{Second}. According to the precedent discussions, 
the observables prepared for a smooth limit from GR to SR admit two 
interpretations; one coming 
from a gauge fixing procedure, the other keeping the full gauge freedom. 
In the gauge fixing approach, which is the one we have been introducing, 
we select a gauge such that the coordinates are Gaussian and comoving with 
a given system of observers; {\sl this choice of coordinates is the 
only effect 
of the presence of the observers in our physical framework}. Now the 
observables must only be gauge invariant
in a restricted sense (under the gauge transformations that preserve the 
system of observers). If we wish to restore
the full diffeomorphism invariance of the observables we must first go back 
to a diffeomorphism invariant description of the observers. This is the 
second interpretation of the observables: 
as long as observers enter the stage, 
the search for observables becomes the search for 
diffeomorphism invariant quantities made up with the dynamical entities
(particles and fields) actively present in the physical scenario {\it plus
the observers' world lines}, that are kinematical entities 
playing a passive role in the physical setting.    

\underline{Third}.
In many physical applications of GR the system of observers is
not of the type discussed here. For instance, in terrestrial laboratory
experiments of the gravitational redshift, observers sitting
in different radial coordinates
can not be synchronized. In fact, this impossibility of 
synchronization is deeply related to the gravitational 
redshift. On the other hand, in experiences of observation of the bending of 
light in a path from the stars passing close to the Sun, observers are 
considered as placed in an 
asymptotic region and so they satisfy our conditions. 
In conclusion, there are different ways to place observers in 
a region of spacetime and not all of them are 
suited for a limiting procedure to SR.


\section{Mathematical versus physical gauge freedom. 
Some gauge fixing surprises.}

Consider theories derived from a variational principle. From the purely 
mathematical perspective, 
infinitesimal symmetries whose parameters are arbitrary functions 
are gauge, whereas infinitesimal symmetries whose parameters are  
(infinitesimal) constants are rigid. This is mathematics. Gauge 
transformations will be associated with and generated by constraints 
in the Dirac sense, whereas rigid symmetries will be associated with and
generated by ordinary constants of motion. From the point
of view \cite{dirac64} of physical determinism, mathematical gauge symmetries 
must correspond
to transformations that do not change the physical state, the reason being 
that gauge transformations can link different solutions of the equations of
motion that satisfy the same set of initial conditions and therefore these
different solutions must be considered physically equivalent. 

So far the mathematical and physical concepts of gauge and rigid symmetries 
are in quite a trivial correspondence. But exceptions abound. The first 
exception has been already commented upon in the introduction: in the process 
of a partial gauge fixing the reduction at the level of the gauge group and
the reduction at the level of the variational principle do not 
necessarily commute. Let us clarify this important 
point by considering again the limiting procedure. 

\subsection{Problems associated with a partial gauge fixing}

We start with a full-fledged general invariant theory described by 
a Lagrangian density   
$${\cal L}(g_{\mu\nu},\phi^a,g_{\mu\nu,\sigma},\phi^a_{,\sigma},...  )\ ,
$$ 
where $\phi^a$ represent the fields other than the metric 
(there can be boundary terms for the
action, in order to guarantee that the variational principle yields the
right equations of motion). ${\cal L}$ is supposed to be a scalar density. 
Gauge freedom for this theory will contain the diffeomorphism invariance
and perhaps other invariances (for instance, in an Einstein-Yang-Mills
theory there will be the gauge invariance associated with  the YM part).

Next, consider the limiting case where gravitational effects can be 
neglected and then choose a gauge fixing that sets 
$g_{\mu\nu}=\eta_{\mu\nu}$ everywhere (or just in the region of spacetime 
we are interested with). The Lagrangian density is reduced to 
${\cal L}_r= {\cal L}_{|_{(g_{\mu\nu}=\eta_{\mu\nu})}}$. Now it is
obvious that, while there remains the residual gauge freedom associated
with the
diffeomorphisms that keep invariant the form of the metric 
$\eta_{\mu\nu}$ (that is: the Poincare transformations), this residual
gauge freedom can no longer be retrieved from ${\cal L}_r$ as such 
gauge freedom in the mathematical sense, for the only mathematical gauge
freedom available in ${\cal L}_r$, if any, comes from the other invariances 
that could exist in the original Lagrangian besides diffeomorphism invariance.

Therefore, when one is presented with a Lagrangian like ${\cal L}_r$ in the
framework of SR, there are two ways for its physical interpretation: either 
a) we conceive it as the ---classical--- ultimate description
of the physical system we are dealing with, and in such case the gauge 
freedom will only 
be the one mathematically derived from ${\cal L}_r$; or b) we consider
that any fundamental Lagrangian that neglects gravity is only an
approximation to the correct theory; this correct theory implementing 
general invariance
(There are methods to reintroduce a dynamical metric -or a tetrad- within 
${\cal L}_r$ in order to make it general invariant). In this case, the 
physical gauge freedom available for ${\cal L}_r$ may be larger that
the mathematically derived from it. 
If we think a), then
Poincar\'e group is not gauge. If we think b), Poincar\'e group is the 
residual gauge group that is left after the introduction of the gauge fixing 
$g_{\mu\nu}=\eta_{\mu\nu}$. As long as Poincar\'e
invariant theories are conceived as approximations where the role
of gravity has been deemed irrelevant, the correct interpretation is b),
but the second remark at the end of section 2 shows that the final answer
depends on the way we deal with observers and observables. The two choices,
a) and b), are available, and indeed the choice a) is the most usual in the
framework of SR.

We refer to \cite{Pons:1998tt} for an example where a reduction
process similar to the one presented above is fully accomplished in 
Bianchi-type cosmologies. In that case, 
a partial gauge fixing and the implementation of the Killing symmetries leads 
to a reduced Lagrangian whose mathematically associated gauge 
group is the group of time reparameterizations; instead, from the point 
of view of reducing the gauge group, in addition to time reparameterizations
there remain, as part of the residual gauge group, some special 
three-space diffeomorphisms ---the homogeneity preserving diffeomorphisms
\cite{ashtekarsam91}. Therefore, we are unable to retrieve the full 
residual gauge group from the
information provided by the reduced Lagrangian itself. The gauge group 
mathematically associated with the reduced Lagrangian is a subgroup of
the complete residual gauge group. The physical consequences of this analysis
are immediate, because the number of physical degrees of freedom depends
upon the interpretation we take. 

It is worth mentioning that the reduced Lagrangian might suffer other
drawbacks: constraints may be lost 
in the process of reduction of the variational principle 
(see \cite{Pons:1996ss} \cite{Pons:1998tt}), even though the variables 
involved in these constraints are not eliminated in the reduction process. The
only remedy for this problem is to reintroduce from the outset these lost
constraints as restrictions on the initial conditions.      


\subsection{Problems associated with boundary conditions}

Another source of conflict between the mathematical and the physical concept
of gauge symmetry comes from boundary considerations. Here
we examine a toy model where the ``mathematical'' and the ``physical'' 
concepts of gauge symmetry do not coincide because of this effect. This 
example is inspired in an earlier model introduced in \cite{Marolf:1996kh}.

Consider the following action in  $1+1$ dimensions with fields $K,\, M,\, N$ 
$$
S= \int_{t_1}^{t_2} dx\, L , \qquad 
L = \int_{x_1}^{x_2} dx \,{\cal L} = 
\int_{x_1}^{x_2} dx \, \left(K(x,t) {\dot M}(x,t) - N(x,t) M(x,t)\right)
$$

This model exhibits the gauge symmetry

\bea
\delta K &=& -\epsilon(x,t) \nonumber \\
\delta M &=& 0 \nonumber \\
\delta N &=& {\dot \epsilon(x,t)},
\label{gauge} 
\eea
with $\epsilon$ arbitrary. Under this transformation, 
$$
\delta {\cal L} = {d \over dt}(- \epsilon M).
$$ 
Although the model lacks physical interpretation, we will accept (\ref{gauge})
as the ``physical'' gauge freedom. In principle, 
regarding the boundary conditions for the application of the 
variational principle, one could consider that the
gauge transformations are restricted to satisfy 
$\epsilon(x,t_1)= \epsilon(x,t_2) = 0$. But consider the piece of
trajectory between $t'_1$ and $t'_2$, for $t_1<t'_1<t'_2<t_2$. Now
$\epsilon(x,t'_1)$ and $\epsilon(x,t'_2)$ are free, and yet the transformed
trajectory must be gauge equivalent to the original trajectory. So we
conclude that the variational principle by itself imposes no
restrictions to the boundary values of the arbitrary function 
$\epsilon(x,t)$.

Let us now introduce a boundary condition on the solutions of the 
equations of motion. We choose, inspired in \cite{Marolf:1996kh},
\beq
\int_{x_1}^{x_2} dx\, N(x,t) = constant.
\label{bc}
\eeq

This boundary condition restricts our gauge freedom, for we need
$$
0 = \int_{x_1}^{x_2} dx\, \delta N(x,t) 
= \int_{x_1}^{x_2} dx\,{\dot \epsilon(x,t)}.
$$
Therefore our physical gauge freedom is now (\ref{gauge}) supplemented with 
\beq
\int_{x_1}^{x_2} dx\,{\dot \epsilon(x,t)} = 0 \ ,
\label{restr}
\eeq
or, equivalently,
$$
\int_{x_1}^{x_2} dx\,{ \epsilon(x,t)} = constant. 
$$

Let us now proceed to the canonical analysis of the model.

The momenta, as functions in tangent space, are
\bea
{\hat P_K} &=& {\partial L \over \partial {\dot K}} = 0 \\
{\hat P_M} &=& {\partial L \over \partial {\dot M}} = K \\
{\hat P_N} &=& {\partial L \over \partial {\dot N}} = 0,
\eea
and the Dirac Hamiltonian becomes
\beq
H_D = \int_{x_1}^{x_2} dx\, (NM + \lambda P_K + \mu (P_M - K) + \eta P_N,
\label{hd}
\eeq
with $\lambda, \mu, \eta$ arbitrary functions.

The constraints $P_K$ and $P_M - K$ are second class and can be eliminated
with the introduction of the Dirac bracket. $K$ becomes the canonical
conjugate to $M$, with the equal-time bracket
$$
\{M(x,t),\,K(y,t) \} = \delta(x-y). 
$$
The bracket $\{N(x,t),\,P_N(y,t) \}$ needs more work, because of the boundary
condition (\ref{bc}). Expansion in Fourier modes for the space 
variable tells us that
all the modes for $N(x,t)$ are configuration variables except 
the zero mode, which is a constant according to (\ref{bc}). 
The canonical conjugates of the non-zero modes define the 
canonical conjugate variables. Therefore the spatial Fourier expansion of 
$P_N(y,t)$ gives a similar picture: the zero mode is absent and the
non-zero modes are the variables. Should all the modes be variables, 
the completion formula would give $\{N(x,t),\,P_N(y,t) \}= \delta(x-y)$, 
but we must subtract to it the contribution of the zero modes. We find
$$
\{N(x,t),\,P_N(y,t) \}= \delta(x-y) - {1 \over x_2-x_1},
$$
which is consistent with 
$$
\{\int_{x_1}^{x_2} dx\, N(x,t),\,P_N(y,t) \}=
\{N(x,t),\,\int_{x_1}^{x_2} dy \,P_N(y,t) \}=0
$$

Let us look for the secondary constraints, Under our Dirac bracket, the
Dirac Hamiltonian is just $H_D = \int_{x_1}^{x_2} dx \, (NM + \eta P_N)$. 
Stabilization of the constraint $ P_N = 0$ gives
$$
0={\dot P_N(x,t)} = \{ P_N(x,t),\,H_D \} =
- M(x,t) + {1 \over x_2-x_1} \int_{x_1}^{x_2} dy\, M(y,t) =: -\Psi(x,t)
$$
This constraint $\xi(x,t) = 0$ is equivalent to the equation 
of motion $M'(x,t) = 0$ \footnote{
$M'(x,t)=0$ is a Lagrangian equation of motion because the variation
of the Lagrangian with respect to the variable $N(x,t)$ must take into
account the boundary condition (\ref{bc}). The Lagrangian is only varied
with respect to the non-zero modes of $N(x,t)$ and the associated equation
of motion imposes that the non-zero modes of $M(x,t)$ vanish. 
This is the contents of $M'(x,t)=0$.} 
($M'$ stands for the space derivative of $M$). 
This new constraint is first class, and there are no more constraints.

Let us construct the Noether gauge generator 
\cite{Batlle:1989ek,Pons:2000az} provided by the first class
constraints. The mathematical construction gives
\beq
G_{math} = \int_{x_1}^{x_2} dx\, \left(\epsilon(x,t) \Psi(x,t) 
+ {\dot \epsilon}(x,t) P_N(x,t)\right) \ , 
\label{g}
\eeq
which is subject to the restriction that it must generate transformations 
that respect (\ref{bc}). This is the case indeed because, with 
the Dirac brackets determined before, we find 
\bea
\delta K &=& \{K,\, G_{math} \} = - {\bar \epsilon(x,t)}  \nonumber \\
\delta M &=& \{M,\, G_{math} \} = 0 \nonumber \\
\delta N &=& \{N,\, G_{math} \} = {\dot {\bar \epsilon}}(x,t) \ ,
\eea
(to be compared with (\ref{gauge}) and (\ref{restr})) 
where we have defined
$$
{\bar \epsilon(x,t) } = 
\epsilon(x,t) - {1 \over x_2-x_1} \int_{x_1}^{x_2} dy\,\epsilon(y,t).
$$
Notice that ${\bar \epsilon}$ satisfies 
\beq
\int_{x_1}^{x_2} dy \,{\bar \epsilon(y,t) }=0 \ ,
\label{restrbar}
\eeq
which is in fact more restrictive than (\ref{restr}). The most 
general function $\epsilon(x,t)$ satisfying (\ref{restr}) is  
\beq
\epsilon(x,t)={\bar \epsilon(x,t)} + a,
\label{decomp}
\eeq 
with ${\bar \epsilon(x,t)}$ satisfying (\ref{restrbar}) and 
$a$ an arbitrary constant 

(Note that $a$ is retrieved from $\epsilon(x,t)$ satisfying (\ref{restr}) as
$$
a = {1 \over x_2-x_1} \int_{x_1}^{x_2} dx\,\epsilon(x,t) \ .)
$$

$G_{math}$ is equivalently written as
\beq
G_{math}[\bar \epsilon] = \int_{x_1}^{x_2} dx\, ({\bar\epsilon}  M + 
{\dot{\bar \epsilon}} P_N). 
\label{g2}
\eeq
(Equation (\ref{restrbar}) makes $G_{math}[\bar \epsilon]$ 
to vanish on-shell)
Notice that the complete gauge generator of the symmetry (\ref{gauge}), with
the restriction (\ref{restr})
is not described by (\ref{g2}). Indeed the physical gauge generator is 
\beq
G_{phys}[\epsilon] = \int_{x_1}^{x_2} dx\, ({\epsilon}  M 
+ {\dot{\epsilon}} P_N), 
\label{gtrue}
\eeq
with ${\epsilon}$ satisfying (\ref{restr}). Using the decomposition
(\ref{decomp}) we find
\beq
G_{phys}[\epsilon] =G_{math}[\bar \epsilon] +  
\int_{x_1}^{x_2} dx\, a  M(x,t).
\label{true}
\eeq

The first term in (\ref{true}), $G[\bar \epsilon]$, is made up 
with first class constraints, because (\ref{g}) and (\ref{g2}) are
equivalent when $\epsilon$ satisfies (\ref{restr});
the second term, $\int_{x_1}^{x_2} dx\, a  M(x,t)$, is made up with the
constant of motion $M(x,t)$. So we see that in this model the gauge symmetry 
(in the physical sense) is generated both by constraints and by constants 
of motion.

\vspace{5mm}

Now we can proceed to a gauge fixing. We fix $N(x,t)$ to be a given
configuration $N_0(x,t)$ satisfying (\ref{bc}). The reduced
Lagrangian becomes
$$
{\cal L}_r = K{\dot M} - N_0M.
$$

The residual gauge freedom is generated by the second term in (\ref{true}).
It gives the transformations 
$$
\delta K = - a, \quad \delta M = 0, 
$$
which is a --physical-- gauge transformation generated by a constant of 
motion. Notice, though, that there is no --mathematical-- gauge freedom 
stemming directly from
${\cal L}_r$ because its two canonical constraints are second class.

\vspace{5mm}

\section{Asymptotically flat spaces: Poincar\'e group as gauge group}

The study of asymptotically flat spaces in GR has led to a distinction
between what is physically gauge and what is not that has become 
commonplace \cite{faddeev82}. Briefly, diffeomorphisms that become
the identity at space infinity are taken as gauge, while diffeomorphisms 
that ``move the 
space boundary'' are considered to have a non gauge part that can be 
identified with the Poincar\'e group. Let us be more explicit.

Here we follow Faddeev approach \cite{faddeev82}. Asymptotically flat 
spaces correspond  
to physical situations where the gravitating masses and matter fields
at finite times are effectively concentrated in a finite region of space.
In a topologically simple spacetime manifold whose points can be uniquely 
parametrized by four coordinates $x^\mu, \, -\infty < x^\mu < \infty $, 
a system of coordinates admissible in order to define an asymptotically 
flat space is such that, in the limit $r \rightarrow \infty $ 
($ r^2 = (x^1)^2+(x^2)^2+(x^3)^2$) for finite time $t = x^0$, the metric 
components satisfy the asymptotic conditions 
\beq
g_{\mu\nu} = \eta_{\mu\nu} + O({1 \over r}), \quad
\partial_\sigma g_{\mu\nu} = O({1 \over r^2}), \quad
\Gamma_{\mu\nu}^\sigma  = O({1 \over r^2}).
\label{asym}
\eeq  

A first partial gauge fixing is clearly under way: we will only 
accept changes of coordinates that preserve (\ref{asym}). 
For an infinitesimal
change $x^\mu \rightarrow x'^\mu = x^\mu - \epsilon^\mu(x)$, this means
that $\epsilon^\mu$ must asymptotically behave as
\bea
\epsilon^\mu &=& \omega^\mu_\nu x^\nu + a^\mu + O({1 \over r}), 
\nonumber \\
\partial_\nu\epsilon^\mu &=& \omega^\mu_\nu + O({1 \over r^2})
\nonumber \\
\partial_\nu \partial_\sigma\epsilon^\mu &=& O({1 \over r^{2 + \alpha}}),
\, \alpha > 0.
\label{ep-restr}
\eea
($\omega^\mu_{\ \nu}$ is an infinitesimal Lorentz parameter, 
$\eta_{\rho\mu}\omega^\mu_{\ \nu} + \eta_{\nu\mu}\omega^\mu_{\ \rho} = 0$, 
and $a^\mu$ is an infinitesimal translation of the coordinates)
Under this partial gauge fixing, our remaining diffeomorphisms group $G$ is 
the one generated by the transformations
satisfying (\ref{ep-restr}). This group has a normal subgroup $G_0$ 
generated by
the transformations that become the identity for $r \rightarrow \infty $. 
The factor group is just the Poincar\'e group. It is then usual to 
take the normal subgroup as the physical gauge group of the theory 
whereas the Poincar\'e group, 
is considered as non-gauge.   

Intuitively, this procedure of singling out the Poincar\'e group as 
non-gauge relies on the consideration that the Poincar\'e group in SR 
is never taken as a gauge group, and since in our case, when $r$ goes
to infinity
our spacetime manifold becomes indistinguishable from Minkowski space, this 
Poincar\'e group that ``moves the boundary'' should not be taken as gauge.
On the other hand, the Hamiltonian
density for this theory  \cite{regge74} \cite{faddeev82} 
differs from the Hamiltonian constraint by a space divergence, and the 
computation of the energy becomes an integration over a closed 
two-dimensional surface $S$ in the limit 
$r \rightarrow \infty $:

$$
E = \lim_{r\rightarrow \infty} \int_S d\sigma_i \ (\partial_j g_{ij}
- \partial_i g_{jj}) \ .
$$

Then, keeping the interpretation of Poincar\'e group as non-gauge, 
$E$ is invariant under 
the ``gauge transformations'' generated by $G_0$. Instead, since a Lorentz 
transformation will change the value of $E$, the energy would not be
a gauge invariant concept should Poincar\'e group be taken as gauge. 

Now our considerations on observers in GR and 
SR made in sections 1 and 2 enter the stage: 
as long as a system of observers is not introduced, the Poincar\'e group 
must be taken as part of the gauge group. According to this view, the concept 
of energy is no longer gauge invariant. 
 The rationale of this conclusion can be made even more clear with the
simplest of the examples. Consider in Minkowski spacetime a physical system 
consisting of a single particle. If, strictly, this is the whole 
physical system, there is no room for observers, nor even in SR. We can 
describe the particle motion in a given Cartesian coordinate frame, 
but we can not attach a 
physical sense -as a reference frame related to a system of observers- to that
particular description. 
We are bound to recognize that the energy, or the momentum, of the particle, 
are no longer gauge invariant concepts, for there are no observers to 
which these concepts can refer. Any Minkowski spacetime with a single particle
(the same type of particle), whatever the coordinate description we take 
for its motion, is always the same physical situation. The perhaps 
surprising fact, then, that constants of motion ---not constraints--- 
may generate physical gauge transformations ---though, mathematically, 
they are rigid transformations--- is less surprising after
the discussion of the toy example of subsection 3.2.

We conclude that the common assertion that diffeomorphisms that 
change the space boundary are not gauge transformations (that is, that they 
change the physics) is, face value, too rash an assertion, 
and that it can only make sense after the role of the observers is 
considered.

\section{conclusions}

We have discussed the role of observers in GR in view of the
limiting procedure from GR to SR when gravitational effects can be
ignored. We define a system of privileged observers in GR that keep
good properties of synchronization with their neighbors. The world lines of 
these systems of observers are geodesics and irrotational. 
This system of observers is ideally constituted 
by test bodies that gather information from their surroundings but
do not affect neither each other nor the spacetime manifold. When the 
the limit is taken from GR to SR they will correspond to the standard 
concept of observers used in SR. 
 
We show that, with regard to a physical description through  
a system of observers, two interpretations can be given. In the first, 
a partial gauge fixing is made in order to keep
the coordinates comoving with the observers; then the observables must only be 
invariant under the elements of the diffeomorphism group that preserve 
the system of observers. This is
not in contradiction with the relational concept of observables 
\cite{Rovelli:1991ph} because the readings of the observables 
are referred
to the system of observers. In fact, this consideration leads to the 
second interpretation: when the world lines
of the observers are included in the physical scenario, the same
observables can be described as gauge invariant 
quantities.
  
Our claim is that spacetime gauge symmetry ---diffeomorphism invariance---
is special among all gauge symmetries, 
and what makes it special is the role of the observers.

We also discuss several issues that suggest the relevance of distinguishing,
in some situations,
between a physical concept of gauge transformation and a mathematical one, 
because they do not always coincide. 
Problems that raise this issue include reduction procedures, where we 
show that the reduction 
at the level of the gauge group and the reduction at the level of the 
variational principle do not necessarily commute, and other problems 
originated by 
the setting of boundary conditions. In particular, we discuss the gauge group
for asymptotically flat spaces, and we conclude in this case that the 
Poincar\'e transformations, that change the space boundaries, must be 
interpreted as gauge transformations unless we have placed a system of
observers on which our previous discussion applies.

\vspace{13mm}



\end{document}